\documentclass[10pt, a4paper]{article}

\newcommand{\q}{\hbox{\rlap{$\sqcap$}$\sqcup$}}

\begin{document}

\title{Spacetime geometric structures and the Search for a Quantum Theory of Gravity}

\author{M. Iftime}

\date{}
\maketitle

\begin{abstract}

One of the biggest challenges to theoretical physics of our time is to find a background-independent quantum theory of gravity. 

Today one encounters a profusion of different attempts at quantization, but no fully accepted - or acceptable, theory of quantum gravity. Any such approach requires a response to a question that lies at the heart of this problem. ``How shall we resolve the tension between the background dependence of all hitherto-successful quantum theories, both non-relativistic quantum mechanics and special-relativistic quantum field theory, and the background independence of classical general relativity?'' (see \cite{Stachel2006})

The need for a background-independent quantization procedure forms the starting point of my approach. In this paper I shall present a gauge-natural formulation of general relativity, and provide some insights into the structure of the space of geometries, which plays an important role in the construction of a non-perturbative quantum gravity using a path integral approach, as well as in string theory (see e.g., \cite{Deligne, Loll, Witten})

\end{abstract}



\pagebreak

\section{Historical-Physical Motivation}

The most succesful among all gravitational theories is Einstein's theory of general relativity. In general relativity the field equations \footnote{Einstein empty space-time equations: $Ric(g)=0$} are described in purely geometric terms: the space-time is a fairly smooth manifold $M$ of dim $n=4$ with a normal hyperbolic Riemannian structure -- Lorentzian metric tensor $g$ transforms at each point to the flat Minkowski metric $\eta=diag(1,1,1, -1)$, and its timelike and null geodesics represent the paths of freely falling particles and light rays.

General relativity is a dynamical theory; there is no a priori space-time metric -- in order to define concepts such as causality, time or evolution, one must solve first Einstein's field equations, and then construct a space-time (i.e., single out a Lorentzian metric). 
\begin{quote}
"There is NO Kinematics without Dynamics!" (J. Stachel,\cite{Stachel2002})
\end{quote}
On the other hand, quantum physics can only be formulated in a fixed space-time ( e.g., Minkowski flat space-time).

The search for a theory of quantum gravity is basically the search for a more fundamental theory, which will include quantum field theory as one limit and classical general relativity as another.

First attempts to quantize gravity started very early. In 1916, A. Einstein points out: (see \cite{Einstein}) 

\begin{quote}

''Nevertheless, due to the inneratomic movement of electrons, atoms would have to radiate not only electromagnetic but also gravitational energy, if only in tiny amounts. As this is hardly
true in nature, it appears that quantum theory would have to modify not only Maxwellian electrodynamics but also the new theory of gravitation.''

\end{quote}
There are two main approaches to quantize gravity, following the successful of quantization of the electromagnetic field, and they are: canonical and covariant approaches. In the canonical approach (also called the Wheeler's program), one attempts to use the Hamiltonian formulation of general relativity as a stepping stone to quantization. 

In the covariant approach:
\begin{quote}
"Sacrificing the fusion of gravity and geometry is a moderate price to pay for ushering-in the powerful machinery of perturbative quantum field theory." (S. Weinberg, \cite{Ashtekar_winding}) 
\end{quote}
Approaches to quantum gravity face two kinds of problems: the first kind of questions refer to each individual program, while the second type of questions are conceptualand they address important physical issues that underlie the whole subject. For example, questions about the microscopic structure of space-time, the space-time general covariance and background independence. 

Our previous work on the so-called Einstein ''hole argument'' 
(\cite{Iftime_Stachel2006}, \cite{Iftime2006a}) has led us to an extension of the concepts of ''general covariance'', ''background independence'' and superspace from metric theories (like general relativity) to theories constructed on arbitrary gauge-natural bundles. 

In view of the uncertainty, in which direction the search for such a more fundamental theory of gravitation will take theoretical physics, we generalized the hole argument to non-continuum theories, such as causal set theory based on the principle of maximal permutability. (\cite{Stachel2002, Iftime2006b})

In this paper, I shall use the language of $G$-structures, to  give a gauge-natural formulation of general relativity, focusing on the second type of questions, by recalling some of the long standing issues that any satisfactory quantum theory of gravity should address.

\section{Introduction}

Geometric objects on manifolds form bundle functors. This point of view was introduced by Nijenhuis (1972), who defined the concept of natural bundles.\cite{Nijenhuis}. The concepts of gauge natural bundles and gauge natural operators were introduced by Eck (1981).\cite{Eck}

Examples of geometric objects forming natural bundles are tensor fields, connections, covariant derivatives, etc. Natural bundles are defined through functors that, for each type of geometric object, associate a fiber bundle over each manifold, such that geometric objects are sections of natural bundles. In more detail, a natural bundle (or bundle functor) on manifolds is a covariant functor $\mathtt{F}$ from the category ${\mathcal M f_m}$ of $m$-dimensional differentiable manifolds and local diffeomorphisms into
the category  ${\mathcal FM}$ of fibered manifolds, and fiber-preserving morphisms satisfying the following two properties:

\begin{enumerate}

\item (locality) For every open set $U\subseteq M$, the inclusion map $i:U\hookrightarrow M$ is transformed 
into the inclusion map $Fi:FU=\pi^{-1}_{M}(U)\hookrightarrow FM$

\item (regularity) Smoothly parameterized systems of local diffeomorhisms are transformed into smoothly parameterized systems of fibered local automorphisms.(i.e., if $\phi :T\times M\to M$ is a smooth map such that for all $t\in T$ the maps $\phi_{t}:M\to M$ are local diffeomeorphisms, then the maps $F\phi :T\times FM\to FM$  defined by $F\phi(t,)=F\phi _t$ are smooth.)

\end{enumerate}

Natual bundles admit natural local trivializations that can be canonically constructed from local atlases of the base space over which they are defined. For example, the tangent bundle $(TM\to M)$ is a natural bundle $T:{\mathcal M f_m}\to{\mathcal FM}$ taking a manifold $M$ into the tangent bundle $(TM\to M)$ over $M$, and local manifold diffeomorphisms $f:M\to N$ into bundle isomorphisms $f_{*}:TM\to TN$ ( dim $M$ $=$ dim $N$).

A {\em natural} morphism $\bar{f}:E\to E'$ is a bundle morphism that is the natural lift of a local base diffeomorphism $f:M\to M'$.  A {\em covariant} transformation $\tilde{f}:E\to E$ is a bundle morphism of $E$ that is the lift of a (global) diffeomorphism of $M$. Therefore there is an action of the group $Diff M$ on the value $(E\to M)$ of a natural bundle over $M$, that takes sections into sections. An example of a natural morphism on the tangent bundle is the tangent map: if $f:M\to M$ local diffeomorphism, then $f_{*}:TM\to TN$ is natural, because it maps linearly fibers and projects on $f$. 

However on a natural bundle there are bundle morphisms that are not natural, or covariant. The naturality property is equivalent with the existence of a splitting of the sequence: $\{e\}\to VertAut(E) \stackrel{i}{\hookrightarrow} Aut(E) \stackrel{pr}{\longrightarrow } Diff M\to \{e\}$, where $VertAut(E)$ denotes the group of automorphisms of $(E\to M)$ that projects over the identity $id_{M}:M\to M$, $Aut(E)$ the group of automorphisms of $E\to M$  and $\{e\}$ is the trivial group. (See e.g., \cite{Fatibene, Iftime2006b})

If $F$ is a natural bundle of order $r$( i.e., any two local base diffeormphisms that coincide up to $r$-th order jet prolongation, their natural lift fiber-preserving morphism coincide) then on the value $(E\to M)$ of $F$ over a $m$-manifold $M$, there is a canonical fiber bundle structure with structure group $G_{m}^{r}$(the $r$-order jet group) which acts on the left on the standard fiber $F_{0}R^{m}=p^{-1}(0)$, where $(FR^{m}\stackrel{p}\to R^{m})$. \footnote{For more details see\cite{KMS}.}

Natural bundles themselves can be considered as objects in a category, the category of natural bundles. \footnote{It is possible to construct a whole family of natural bundles defined on a fixed $M$, e.g., tensor bundles.} The objects in this category are natural bundles, and the morphisms are natural transformations. For example, the vector bundle $\Lambda ^{k}T^{*}M\to M$ of exterior forms over $M$ forms a natural functor $\Lambda ^{k}T^{*}$. The property $df^{*}\omega = f^{*}d\omega$ translates into: the exterior derivative $d$ is a natural transformation from the natural bundle $\Lambda ^{k}T^{*}$ into $\Lambda ^{k+1}T^{*}$.\cite{KMS}

A covariant theory\footnote{Some authors define general covariance as what we call covariance, not making a clear distinction between the two.
See \cite{Iftime_Stachel2006, Stachel1993}.} on a manifold $M$ is a physical theory defined on a natural bundle $(E\to M)$ over $M$ such that  the field equations defining the theory are invariant under the action of $Diff M$. In general, the physical field equations can be brought into  to the form of a first-order, quasilinear (i.e., linear in the first order derivatives\cite{GerochPDE}) system of partial differential equations\footnote{Except in some cases with degenerate Lagrangians}. Physical fields ( represented by sections of a natural bundle $(E\to M)$) are subject to field equations that can be deduced from the variation of a first order Lagrangian $L:J^{1}E\to \Lambda^{m}T^{*}M $ (a bundle morphism over $M$) \footnote{Except some cases, for example, for dissipative systems, (e.g., the hydrodynamical equations
with viscosity added) and some unified field theories-- Einstein and Walther Mayer gave up a unified field theory in part because it was not  derivable from a Lagrangian}
In other words, such a physical system can also be defined by the vanishing of a collection of differentiable functions on the first jet space $J^{1}E$ of the configuration space $(E\to M)$ - a variety, which is a closed submanifold $S\in J^{1}E$, solutions of which are represented by sections $\sigma$ of the natural bundle $(E\to M)$ such that $j^{1}\sigma\in S$. The group $Diff M$ of all diffeomorphisms of $M$ acts on the natural bundle $(E\to M)$ of a covariant theory, taking solutions into solutions  (i.e., symmetries of the field equations -- the Lagrangian of the system is invariant under the action of $Diff M$). So, if $\sigma$ is a solution (i.e., $\sigma$ is section of $E\to M$ and $j^{1}\sigma\in S$), then all the pullback sections $\phi^{*}\sigma$ are also solutions ( $j^{1}_{\phi(x)}\phi^{*}\sigma=j^{1}\Phi(j^{1}_{x}\sigma)\in S$ 
for all $x$, where $\Phi:E\to E$ is the covariant lift of the base diffeomorphism $\phi:M\to M$ ). Thus, a covariant theory possesses gauge freedom associated with global diffeomorphisms of the underlying manifold, and one can formulate a version of the hole argument for covariant theories. \cite{Iftime_Stachel2006} 

General Relativity is a covariant field theory; it can be defined on the natural bundle of Lorentz metrics over a $4$-dimensional manifold $M$ with the (second order) Hilbert Lagrangian $L(g)=R\sqrt{|det(g)|}$.  Einstein's field equations, $Ric(g)=0$ possess gauge freedom associated with global diffeomorphisms of the underlying manifold $M$. The symmetry group of the Einstein action is the group $Diff(M)$ of diffeomorphisms on $M$. The group $Diff(M)$ acts by pulling back metrics on $M$: for all $\phi\in Diff(M)$ and $g$ on $M$, the action of $Diff(M)$ defined by $(\phi, g)\longmapsto \phi^{*}(g) $ partitions the space $\mathcal{M}(M)$ of all metrics on a given differentiable manifold $M$ into (disjoint) isometry classes; a physical gravitational field is represented by a point in this orbit space $Geom(M) =\mathcal{M}(M)/Diff(M)$, the space of $4$- geometries (the superspace).\cite{Iftime2006b, Iftime_Stachel2006}

Classical versus quantum dynamics: The success of path integrals 
in describing non-gravitational field theories has led to attempts to describe gravity using path integrals (see, e.g., \cite{Rovelli2001}). But gravity is rather different from the other physical interactions, whose classical description involves fields propagating in spacetime. In general relativity a metric spacetime $(M, g)$ 
emerges as a solution to Einstein's equations, which define the classical dynamics on $Geom(M)$. In \cite{Loll1} a path integral approach to quantum gravity is considered, which is non-perturbative in the sense that it is background independent; a "quantum space-time" is defined using this approach as a solution of a set of non-perturbative quantum equations of motion on a quantum analogue of the space of all geometries. (\cite{Loll1})

\subsection{Gauge-natural bundles and gauge theories}
 
Gauge natural bundles are the framework  for gauge theories, and as we shall see, one can formulate general relativity as a gauge natural theory.

Gauge natural bundles are defined by starting from a fixed principle bundle: Given $G$ a Lie group, a gauge natural bundle on $m$- manifolds with structure group $G$ is a covariant functor $\mathtt{F}$ from the category ${\mathcal P(G)}$ of principal bundles with structure group $G$ and principal morphisms into the category ${\mathcal FM}$ of fibered manifolds and fiber-preserving morphisms satisfying the locality and regularity properties.\cite{Fatibene, KMS} Gauge natural bundles admit local trivializations that are canonically constructed from local trivializations on the principle bundle $(P\to M; G)$. Examples of gauge natural bundles are all associated bundles $(E(F)=(P\times F)/G\to M)$ of the principle bundle $(P\to M; G)$ with fiber type $F$ and the 'twisted' action of $G$ on $(P\times F)$, $(p,u)\cdot g=(p\cdot g, g^{-1}\cdot u)$.
A gauge-natural theory on a principal bundle $(P\to M; G)$ is a physical theory defined on an gauge natural bundle $(E\to M)$ over $(P\to M; G)$ such that the Langrangian of the physical theory is invariant under the action of the group $Aut(P)$ of principal bundle automorphisms of $(P\to M; G)$ .
The group $Aut(P)$ acts on the associated bundles, transforming sections into sections, and solutions into solutions. Each $\phi_{P}$ induces an automorphism $\phi_{E}:(P\times F)/G\to(\phi_{P}(P)\times F)/G$ of the associated bundle, but $Diff M$ does not ( because there may exists diffeomorphisms of $M$ that are not the projections of a principle automorphism $\phi_{P}:P\to P$)

Because of the gauge freedom, gauge natural theories can only determine a class of gauge-related solutions, and not a representative of the class (single solution). Therefore, a gauge natural theory possesses gauge freedom associated with $Aut(P)$. (see \cite{Iftime2006b}
for a version of hole argument for gauge natural theories.)

Yang-Mills theory is a gauge theory of principal connections on a principal bundle $(P\to M; G)$. Being $G$-equivariant, principal connections on $P$ i.e., gauge potentials, they are identified with global sections of the gauge natural bundle $E = J^{1}P/G$ over $M$, so it is a gauge natural theory. Gauge transformations in Yang Mills theory are vertical automorphisms, i.e. principal automorphisms $\phi_{P}\in Aut(P)$ that project over the identity base diffeomorphism $id_{M}:M\to M$. 
They form a subgroup $Gau(P)$ of $Aut(P)$, the vertical (pure) gauge group .
Yang Mills theory can only determine a class of gauge-related solutions, and not a representative of this class, so it possesses gauge freedom associated with pure gauge group $Gau(P)$.

\section{General relativity as a gauge natural theory and the space of geometries}\label{gr}

General relativity is a natural covariant theory, but it can be also constructed as a gauge-natural theory defined on the principal linear frame bundle $(LM\to M)$. Because the principal linear frame bundle is actually a natural bundle (see e.g., \cite{Iftime2006b}) the two formulations  of general relativity are consistent. The group $Diff M$ of diffeomorphisms acts on $LM$: a spacetime diffeomorphism lifts uniquely to an affine transformation of the basis vectors at each point of the frame bundle. In the case of general relativity, one may say that principal automorphisms are "soldered" to base diffeomorphisms.

There are numerous papers in the literature discussing the problem of including gravity in the universal description of fundamental interactions described by gauge theories. (see e.g., \cite{Utiyama, Hehl, Sardanashvily})
 
One of the main advantages for our formulation of general relativity as a gauge natural theory is that it reveals new insights in the study of the space of all geometries $Geom(M)$ on a given space $M$, which plays an important role in string theory (when $dim M = 2$) as well as in a non-perturbative path integral formulations to quantum gravity ( for $dim M = 4$)\cite{Hawking, Loll}.

Let $M$ be a $4$-dimensional differentiable manifold.
An equivalent formulation of the equivalence principle (see e.g., \cite{Wald}) states that at any local region in spacetime it is possible to formulate the equations governing physical laws such that the effect of gravitation can be neglected. 

The equivalent principle assures the existence of a Lorentz metric on $M$.In bundle terms, implies that the linear frame bundle $LM$ is reducible to the Lorentz subgroup $SO(1,3)$. (see e.g.,\cite{Sardanashvily})

A spacetime structure $E$ on $M$ is a reduced $SO(1,3)$-subbundle of the linear frame bundle $LM$ and we have the following proposition:\\

\textit{Proposition 1}

There is a 1:1 correspondence between the $SO(1,3)$-reduced subbundles of $LM$ and the tetrad gravitational fields represented by global sections of the associated fiber bundle $E(GL(4)/SO(1,3))$ defined as the fiber bundle $\displaystyle(LM/SO(1,3)\stackrel{\pi}{\longrightarrow}M)$
of orbits under the action of $SO(1,3)$ on the linear frames on $M$, with  the standard fiber $GL(4)/SO(1,3)$.\\

\textit{Proof:}\\
Indeed, one can see this immediately from the following sequence of fiber bundles:
$LM\stackrel{p_{1}}{\longrightarrow}LM/SO(1,3)\stackrel{\pi}{\longrightarrow}M $.
One obtains a $SO(1,3)$-subbundle $E\subset LM$, $E=p^{*}_{1}(\sigma(M))$ by pulling back $\sigma(M)\subset LM/SO(1,3)$ of a section $\sigma:M\to LM/SO(1,3)$ by the principal projection map $p_1$.
One can also view a $SO(1,3)$-reduction $E$ as the inverse image of the identity coset in $GL(4)/SO(1,3)$.
So, a spacetime structure $E$ on $M$ can be defined as a pair $(LM, \sigma)$, where $\sigma$ is a global section of$(LM/SO(1,3)\stackrel{\pi}{\longrightarrow}M)$. Such a global section specifies a class of Lorentz transformations-related linear frames at each point in $M$.$\q$\\

Two reduced $SO(1,3)$-subbundles of $LM$ are said to be equivalent if they are isomorphic as $SO(1,3)$ bundles.
 
Let $\Phi\in Aut(LM)$ be a principal automorphism of the linear frame bundle. Then $\Phi$ also acts as an automorphism on the associated bundle $(LM/SO(1,3)\stackrel{\pi}{\longrightarrow}M)$. This gives an action of $Aut(LM)$ on the set of all reduced $SO(1,3)$-subbundle of $LM$. Moreover, one obtains an action of $Diff(M)$ on on the set of all reduced $SO(1,3)$-subbundle of $LM$.(The linear frame bundle $(LM\to M; GL(4))$ is actually a natural bundle, so any base diffeomorphism $\phi:M\to M$ lifts to a (unique) principal bundle automorphism $\Phi\in Aut(LM)$).

Two reduced $SO(1,3)$-subbundles $E_1$ and $E_2$ of $LM$ are equivalent if and only if there exists an automorphism $\Phi\in Aut(LM)$ such that $\Phi(E_{1})=E_{2}$. This implies that $E_1$ and $E_2$ are equivalent if there exists a diffeomorphism $\phi:M\to M$ such that the two global sections $\sigma_{1}$ and  $\sigma_{1}$ ( where $E_{i}=p^{*}_{1}(\sigma_{i}(M))$, for $i = 1,2$ ) are diffeomorphically related.\\

\textit{Proposition 2}

There is a 1:1 correspondence between
\{ Isomorphism classes of $SO(1,3)$-reduced subbundles of $LM$\}  and \{ Diffeomorphically related global section of the associated fiber bundle  $(LM/SO(1,3)\stackrel{\pi}{\longrightarrow}M)$ \}.\\

On the other hand, let $\mathcal{M}(M)$ be the space of all Lorentz metrics on $M$. The group $Diff(M)$ of diffeomorphisms on $M$ acts as a transformation group on $\mathcal{M}(M)$ by pulling-back metrics on $M$: for all $\phi\in Diff(M)$ and $g\in \mathcal{M}(M)$ the action map is defined by $(\phi, g)\longmapsto \phi^{*}(g) $.  The action of $Diff(M)$ on $\mathcal{M}(M)$ partitions $\mathcal{M}(M)$ into (disjoint) isometry classes of metrics, and defines an equivalence relation on $\mathcal{M}(M)$.
The superspace, the space $Geom(M)=\mathcal{M}(M)/Diff(M)$ of all isometry classes of metrics on $M$ is a moduli space of $4$- geometries on $M$.

If we assume now that the Lorentz metrics are subject to field equations, then\\

\textit{Proposition 3}

$Geom(M)$ is the classification space of diffeomorphically related Lorentz metrics subject to Einstein field equations with respect to $Diff M$: each point corresponds to one and only one physical gravitational field. \cite{Iftime2006a} \\

A section of $(LM/SO(1,3)\stackrel{\pi}{\longrightarrow}M)$ determines uniquely a Lorentz metric on $M$ \footnote{$LM/SO(1,3)$ forms a 2-fold covering space of the bundle of all Lorentz metrics on $M$. The restricted Lorentz group $SO^{+}(1,3)$ preserve both the orientation and the direction of time ($+$).}, so by applying the above proposition it implies the following:\\

\textit{Proposition 4}

There is a 1:1 correspondence between 
\{ Points of the moduli space of $4$- geometries $Geom(M)$ on a given $M$\}
and \{ Isomorphism classes of $SO(1,3)$-reduced subbundles of $(LM\to M)$\}.\\

It implies that one can study the structure of the space $Geom(M)$ of $4$-geometries by analysing the structure of the moduli space $\mathcal{M}_{SO(1,3)}$ of all isomorphism classes of $SO(1,3)$ reduced principle subbundles of the linear bundle. 

For lower dimensional manifolds there is a classical result of Grothendieck\cite{Grothendieck}, which shows that such moduli space has an algebraic sheme structure defined in a natural way.

\section{Some remarks on background dependence and independence}

In realistic models of the world, several physical systems coexist on the underlying spacetime manifold $M$, and they are interacting with each other kinematically or dynamically.\footnote{However we do not assume yet a metrics or any other geometrical structures on $M$ and consider only differentiable theories, although some results can be extended to discrete theories.\cite{Iftime2006a}} We denote by $(E_{(X,Y)}\stackrel{p_{X,Y}}{\longrightarrow} M)$ the fiber bundle, with fibers consisting of two fields $X$ and $Y$. If  $Y$ is a background field for $X$ this implies that
we can "forget" the $X$ fields by taking the projection over the space $E_{Y}$ of $Y$ fields, and deal with a quotient bundle:
$E_{(X,Y)}\stackrel{p_{X}}{\longrightarrow}E_{Y}\stackrel{p_{Y}}{\longrightarrow} M$  such that $p_{Y}\circ p_{X}=p_{(X,Y)}$. (see \cite{GerochPDE})
Fixing a background field $Y$ i.e., given a section (solution) $\sigma_{Y}$ of $(E_{Y}\stackrel{p_{Y}}{\longrightarrow} M)$, then the configuration bundle of $X$ fields with the fixed background field $Y$ is defined as the subbundle $\hat {E}^{Y} = p_{X}^{-1}(\sigma_{Y}(E_{Y}))$ 
of $(E_{(X,Y)}\stackrel{p_{X,Y}}{\longrightarrow} M)$ obtained by pulling back $(E_{Y}\stackrel{p_{Y}}{\longrightarrow} M)$ by $\sigma_{Y}$.

A physical theory that contains background fields is called
background-dependent. Examples of background dependent theories are all special-relativistic field theories \footnote{The Minkowski metric (and the compatible Levi-Civita symmetric inertial connection) the represents the background structure of these theories} and theories in a spacetime with a given background metric or other geometric structure. For example  Einstein-Maxwell equations in a fixed background spacetime. In this case the spacetime metric $g$ represents the kinematical background field for the electromagnetic field $F$, and one must solves the empty-space (homogeneous) Einstein equations $Ric(g)=0$ first, then solve the Maxwell field equations $F^{ab}_{;b}=0$ and $F_{[ab; c]}=0$  with that background spacetime\footnote{The covariant derivatives are with respect the Levi-Civita connection compatible with the Lorentz metric $g$}, and then take the stress-energy tensor $T(g,F)$ for that Maxwell field $F$ in the background space-time $g$, treat it as a small perturbation and insert it on the right hand side of the inhomogeneous Einstein equations, $Ein(g)=Ric(g) -\frac{1}{2}gR(g)=T(g, F))$. \footnote{Linearizing them around the original background metric; and calculate a small correction to the original metric, one can keep looping back and forth in this way, hopefully getting better and better approximations to a solution to the (dynamically) coupled Einstein-Maxwell system.(see below a more detailed discussion on this example)
One could call it a "test field" approximation by analogy with a "test
particle" approximation, where one solves for the geodesic motion of a
particle in a given backgroud space-time, neglecting the fact that it
has active mass that will affect the metric.}

A background independent theory is a physical theory defined on a manifold $M$ endowed with no background structure. Examples of background independent theories are all general-relativistic gravitational theories, satisfying either the (homogeneus) Einstein empty-spacetime equations or a set of covariant coupled equations e.g., (inhomogeneous) Einstein-Maxwell or Einstein-Yang Mills. \footnote{Background independence seems to be an important requirement for a corresponding quantum theory of gravitation\cite{Green, Iftime_Stachel2006}} In the case of coupled Einstein-Maxwell or Einstein-Yang Mills equations, the spacetime structure and the source (electromagnetic or Yang Mills) fields constitute a dynamical system: the equations of which can only be solved together.\footnote{For more physical examples see \cite{Geroch, GerochPDE}}. The combined system forms therefore a covariant theory, and it can be modelled on a natural bundle. In general the configuration space for the combined system is a natural bundle $(E_{(X,Y)}\stackrel{p_{X,Y}}{\longrightarrow} M)$, where the fibers consists of all possible values of two fields $X$ and $Y$, that are dynamical coupled. In cases when one of the dynamically coupled fields is a gauge field, one must then restrict to a smaller category of manifolds in order to treat those fields as natural objects, such that there is an action of $Diff M$ on the configuration bundle of the combined system. \cite{Fatibene, Iftime2006b}

\section{Some conclusions}

There are many advantages of interpreting geometric objects as bundle functors. First, one obtains a well-defined mathematical concept, and 
in some cases it is possible to classify and determine all natural operators of a given type.\cite{KMS} Based on this functorial approach, certain geometric constructions can be generalized to other category e.g., from differentiable to discrete, going from covariant to permutable theories by applying a forgetful functor. See \cite{Iftime2006b} for a generalization of the hole argument to permutable theories. 

Functors are useful in keeping track of the relationships between local and global data, and this approach seems very useful in field theory. One can view such local fields as two local expression of a global section.

Our formulation in terms of bundle functors, led us to obtain new insights in the structure of the space of all geometries on a given manifold. A better understanding of this space is important e.g., in string theory and quantum gravity (one is not really looking for a single "quantum space-time", but rather a quantum treatment of space-times). 

In string theory one starts with $1$-dimensional objects that generate $2$-dimensional histories in spacetime, the worldsheets. The theory describing the worldsheet is a sort of $2$D gravitational theory -- here one extremalizes the area of a timelike worldsheet, using the induced $2$-dimensional metric to define the area. (see e.g., \cite{Deligne, Stachel})

Although little is known regarding the existence
and nature of an underlying continuum theory of quantum gravity in  dimension $D = 4$, there are already a number of interesting results in 
$D = 2$ and $D = 3$, where continuum limits have been found. In quantum gravity, one sums over all $4$-dimensional space-time geometries with two $3$-dimensional boundaries which match the initial and final conditions. In practice, calculating the probability amplitude for a transition between the two $3$-dimensional boundaries is a very difficult and an approximation has to be used. (see e.g., the Hartle-Hawking ``No Boundary Proposal''\cite{Hawking}, discrete approaches to quantum gravity in $4$-dimensions \cite{Loll, Loll1}, and symmetry reductions of $4$-dimensional gravity theory \cite{Niedermaier}).

\vspace{15 pt}

Author's address:

\vspace{10 pt}

M. D. Iftime

MCPHS, Boston University

179 Longwood Avenue, Boston, MA, 02115, USA

E-mail: mihaela.iftime@mcphs.edu

\end{document}